
\raggedbottom
\hoffset=0cm\voffset=0cm
\magnification=\magstep1
\baselineskip=15pt
\def\itm#1{\noindent\item{(#1)}}
\def\ref#1{\noindent\item{[#1.]}}
\def\spa#1{\noindent\item {#1}}
\vskip 9cm
\hfill {\tenrm NUS/HEP/94203}
\par\noindent
\hfill {\tenrm hep-th/9408001}
\vskip 2cm
\centerline {\bf Realizations of the $q$-Heisenberg and
$q$-Virasoro Algebras}
\vskip .7cm
{\centerline {C.H. Oh ${}^{\dag}$ and K. Singh }}
\vskip .5cm
{\sl {\centerline {Department of Physics}}}
{\sl {\centerline {Faculty of Science}}}
{\sl {\centerline {National University of Singapore}}}
{\sl {\centerline {Lower Kent Ridge, Singapore 0511}}}
{\sl {\centerline {Republic of Singapore}}}
\vskip 2cm
{\bf {\centerline {Abstract}}}
\vskip .5cm

We give  field theoretic realizations of both the $q$-Heisenberg
and the $q$-Virasoro algebra. In particular,
we obtain the operator product expansions among the current
and the energy momentum tensor obtained using the Sugawara
construction.
\vskip 6truecm
\hrule width6cm
\vskip .3truecm
{\sevenrm \dag E-mail: PHYOHCH@NUSVM.NUS.SG}
\vfil
\eject
\par
Quantum algebras or more precisely quantized universal enveloping
algebras first appeared in connection with the study of the inverse
scattering problem [1]. Subsequently it was shown that these
algebras are also deeply rooted in other areas such as exactly
soluble statistical models [2], factorizable $S$-matrix theory [3]
and conformal field theory [4]. Mathematically, these algebras are
Hopf algebras which are non cocommutative.
These can be compared with the
classical universal enveloping algebras which can be endowed with
cocommutative Hopf structures.
In this regard quantum algebras appear as natural generalizations of
the usual Lie algebras.
\par
Lately there has been a lot of interest in the $q$ deformation of
the Virasoro algebra [6-9]. By generalizing a differential realization of
${\rm su}_{q}(1,1)$, Curtright and Zachos (CZ) [6] obtained a
$q$-analoque of the centerless Virasoro algebra. Its central extension
was later furnished by Aizawa and Sato [7]. However, to date, the
existence of a Hopf
structure for this algebra remains an open question.
\par
More recently Chaichian and Pre${\rm {\check s}}$najder [8] proposed a
different version of the $q$-Virasoro algebra by carrying out a
Sugawara construction on a $q$-analogue of an infinite dimensional
Heisenberg algebra $({\rm H}_q(\infty))$. They also showed that in a
unitary representation, this algebra possesses a primitive
(cocommutative) Hopf structure.
\par
In this paper we realize both the $q$-Heisenberg as well as
the $q$-Virasoro algebra using field operators. In particular,
we obtain the operator product expansions (OPE's) of these field
operators using some of the standard techniques of conformal field
theory. The central term for the $q$-Virasoro algebra
obtained via the realization is shown to differ slightly from the
one given in ref.[8]. It is further shown that our expression leads
to the standard case in the $q\to 1$ limit.
\footnote\dag{Contrary to the claims of ref.[8], the central term
presented there does not have the correct limit.}
\+{~}\hfill\cr
We begin by summarizing some results of ref.[8] that will be used later.
The algebra of ${\rm H}_q(\infty)$ is based on the one dimensional
Heisenberg algebra which is a bona fide Hopf algebra with relations
[10]
\footnote\ddag{It is worth noting that this algebra differs from
those given in refs.[11].}
$$\eqalign{&[a,a^{\dag}]\equiv aa^{\dag}-a^{\dag}a =
{{{\rm sinh}(\epsilon H/2)}\over{\epsilon/2}} \cr
&[H,a]=0,~~~~~[H,a^{\dag}]=0}\eqno(1)$$
with $q=e^{\epsilon}$ as the deformation parameter. Here $a$ and
$a^{\dag}$ can be regarded as the annhilation and creation
operators repectively. The algebra of ${\rm H}_q(\infty)$ is
obtained by considering an infinite collection of these operators
labelled as
$$a_{n},~~a_{-n}=a_{n}^{\dag}~~~~~n=1,2,...,$$
with commutation relations
$$\eqalign{[a_m,a_n] &=\omega_{mn}\cr
[H,a_m]&=0}\eqno(2)$$
where
$$\omega_{mn}={1\over{\epsilon}}{\rm sinh}(m\epsilon H)
\delta_{m+n,0}.\eqno(3)$$
The associative algebra generated by $\{{\bf 1},H,a_n\}_{n\in {\rm
Z}}$ with the above relations can be endowed with Hopf structure by
defining the following:
\itm i co-product:
$$\eqalign{\Delta a_n =q^{-\vert n \vert H/2}\otimes a_n &+
a_n\otimes q^{\vert n \vert H/2},\cr
        \Delta H = {\bf{1}} \otimes H + H \otimes {\bf 1}&,~~
        \Delta {\bf 1}={\bf 1}\otimes {\bf 1};}\eqno(4a)$$
\itm {ii} co-unit:
$$\epsilon (a_n)=0,~~\epsilon (H)=0,~~\epsilon({\bf 1})={\bf 1};
\eqno(4b)$$
\itm{iii} antipode:
$$S(a_n)=-a_n,~~S(H)=-H,~~S({\bf 1})={\bf 1}.\eqno(4c)$$
\par
The $q$-deformed Virasoro generators in the Sugawara construction
read as
$$L_{m}^{\alpha}={1\over 2}\sum_{k,n}{\rm cosh}({{k-n}\over
{2}}\epsilon \alpha H) :a_{k} a_{n}:\delta_{k+n,m}.\eqno(5)$$
The normal ordering prescription here is taken as
$$:a_{k}a_{n}:=a_{k}a_{n}-\theta (k)\omega_{kn}\eqno(6)$$
where $\omega_{kn}$ is defined in (3) and $\theta (k)=1$ or 0 for
$k$ positive or negative respectively. It is worth noting that
$q$-Virasoro  generator carries an additional integer-valued index
which is required for the commutators between the generators to
close. The commutation relations furnished in ref.[8] are given by
$$\eqalign{[L_{m}^{\alpha},L_{n}^{\beta}]=
&{~}{1\over{2\epsilon}}{\rm sinh}({{m-n-n\alpha +m\beta}
\over{2}}\epsilon H)L_{m+n}^{\alpha + \beta + 1}\cr
&+{1\over{2\epsilon}}{\rm sinh}({{m-n+n\alpha -m\beta}
\over{2}}\epsilon H)L_{m+n}^{-\alpha - \beta + 1}\cr
&+{1\over{2\epsilon}}{\rm sinh}({{m-n+n\alpha +m\beta}
\over{2}}\epsilon H)L_{m+n}^{\alpha - \beta - 1}\cr
&+{1\over{2\epsilon}}{\rm sinh}({{m-n-n\alpha -m\beta}
\over{2}}\epsilon H)L_{m+n}^{-\alpha + \beta - 1}\cr
&+{1\over{16\epsilon ^{2}}}(C_{m-1}^{\alpha ,\beta}
+C_{m-1}^{\alpha ,-\beta}) \delta_{m+n,0},}\eqno(7a)$$
with
$$\eqalign{C_{m}^{\alpha , \beta} = &{~}{{{\rm sinh}({{\alpha
+\beta + 1}\over {2}} m\epsilon H)}\over{{{\rm sinh}({{\alpha
+\beta + 1}\over {2}} \epsilon H})}}
-2{\rm cosh}(m\epsilon H + \epsilon H){{{\rm sinh}({{\alpha
+\beta}\over {2}} m\epsilon H)}\over{{{\rm sinh}({{\alpha +\beta
}\over {2}} \epsilon H})}}\cr
&+ {{{\rm sinh}({{\alpha +\beta
- 1}\over {2}} m\epsilon H)}\over{{{\rm sinh}({{\alpha +\beta -
1}\over {2}} \epsilon H})}}.}
\eqno(7b)$$
Here a few remarks are in order. Firstly the $q\to 1$ limit yields
$$[L_{m}^{\alpha},L_{n}^{\beta}]\to [L_{m},L_{n}] =
(m-n)HL_{m+n}-{{1}\over{96}}m(m-1)(11m+2)H^2 \delta_{m+n,0}
\eqno(8)$$
which shows that although the operator part i.e. terms involving the
generators reduce to the usual expression, the central term does not.
Secondly the Hopf structure that can be written down for the
algebra is obvious only for the case when $H$ is a constant or when
a unitary irreducible representation is chosen. In this case,
the algebra becomes an
infinite dimensional Lie algebra and this can be endowed with a
primitive (cocommutative) Hopf structure. In general however, with
$H$ non-trivial in (7), the existence of a Hopf
structure has not be shown. Even if it does exist, as pointed out
in ref.[8], it would most likely be non-trivial and complicated.
\+{~}\hfill\cr
\par
To give a field-theoretic realization of an algebra, one must
essentially furnish the operator product expansion (OPE) between
the appropriate field operators. In fact only the singular part of
this expansion is essential as it embodies all the relevant
information about the algebra. For instance, the usual Heisenberg
algebra can be obtained from the singular portion of the OPE between
two currents:
$$J(z)J(w)={{1}\over {(z-w)^2}}+{\rm regular~terms}.\eqno(9)$$
To obtain a $q$-analogue of this OPE, we begin by defining the
current in the usual way,
$$J(z)=\sum_{m=-\infty}^{\infty}a_m z^{-m-1}\eqno(10)$$
with the operators $\{a_m\}$ satisfying (2) instead of the usual
Heisenberg algebra. Here we will restrict ourselves to the case of
a unitary representation in which $H=1$. To simplify the notation we
write
$$[a_m,a_n]=\kappa~[m]~\delta_{m+n,0}\eqno(11)$$
where
$$[x]\equiv{{q^{m}-q^{-m}}\over {q-q^{-1}}}~~~~{\rm and}~~~~
\kappa = {{1}\over {\epsilon}}{\rm sinh}(\epsilon).$$
Then by using (10) and (11) we have for $\vert w \vert <\vert z\vert$,
$$\eqalign{J(z)J(w) &= \sum_{m,n}a_{m}a_{n}z^{-m-1}w^{-n-1}\cr
                   &= :J(z)J(w):+ {{1}\over {zw}}\sum_{m>0}\kappa
                      ~[m]~({{w}\over {z}})^{m}\cr
                   &= :J(z)J(w):+{{\kappa}\over {(z-w)_{q}^{2}}}}
\eqno(12)$$
where $(z-w)_{q}^{2}=(z-wq^{-1})(z-wq)$. It is interesting to note
that the poles are located at two points, $\{wq^{-1},wq\}$
and both are of order 1. This differs from the standard case where
there is a single pole of order 2 at $z=w$.
However, in the $q\to 1$
limit these poles coalesce to form a pole of order 2 and
expression (12) reduces to (9). The situation here is quite similar
to that of refs.[7] and [12] where a realization for the CZ algebra
[6] also leads to such degeneracies in the poles.
\par
Before proceeding further, it is instructive to check whether the
above $q$-OPE leads to the $q$-Heisenberg algebra as required. To
this end, we first invert (10) to give
$$a_m=\oint_{C}{{dz}\over {2\pi i}}J(z)z^m\eqno(13)$$
where the contour $C$ is taken as $\vert z\vert ={\rm constant}$.
Here we use the usual prescription of radial quantization of
standard conformal field theory in which different `times'
correspond to concentric circles of different radii. In this
context, time-ordering is replaced by that of radial-ordering:
$$R(A(z)B(w))=\cases { A(z)B(w)~~~& $\vert z\vert > \vert w\vert$\cr
B(w)A(z)& $\vert z\vert < \vert w\vert.$\cr}\eqno(14)$$
Then by using the standard procedure for computing an `equal-time'
commutator [13], we have
$$\eqalign {[a_m,a_n] &= [\oint_{C} {dz\over {2 \pi i}}J(z)z^m,
\oint_{C} {{dw}\over {2 \pi i}}J(w)w^n]\cr
&= \oint_{C}{{dw}\over {2 \pi i}}
(\oint _{\vert z\vert > \vert w\vert}-
\oint _{\vert z\vert < \vert w\vert}){dz\over {2 \pi i}}z^{m}w^{n}
R(J(z)J(w))\cr
&=\oint_{C}{{dw}\over {2 \pi i}}
\oint_{C_{P}}{dz\over {2 \pi i}}z^{m}w^{n}R(J(z)J(w))}\eqno(15)$$
where the integral over $z$ is taken around all the poles in the OPE of
$J(z)J(w)$. Now the above procedure only makes sense if we
assume that the singularities of $J(z)J(w)$ are located on the
$\vert z\vert =\vert w\vert$ contour since otherwise these poles will
not make any contribution to the integral. For the $q$-OPE (12) this
requires that  $q$ be a pure phase (or $\vert q\vert = 1$).
Consequently by substituting (12) into (15) and using the fact
$$\oint_{C_P}{{dz}\over {2\pi i}}{{z^m}\over {(z-w)_{q}^{2}}}=
\partial_{z}^{q}z^m\vert_{z=w}=[m]w^{m-1}\eqno(16)$$
we obtain (11).
\par
In analogy to the standard case, we now define the energy-momentum
tensor as
$$T^{\alpha}(z)=\sum_{m}L_{m}^{\alpha}z^{-m-2}\eqno(17)$$
where the index $\alpha$ also appears on $T$ by virtue of its presence
on $L$.
\par
To be consistent with (5), the corresponding Sugawara construction
for $T^{\alpha}(z)$ in terms of the currents reads as
$$T^{\alpha}(z)={{1}\over {4}}:J(zq^{\alpha /2})J(zq^{-\alpha /2}):
+ {{1}\over {4}}:J(zq^{-\alpha /2})J(zq^{\alpha /2}):.\eqno(18)$$
It is worth noting that
the second term on the right hand side is identical to the first
with $q$ and $q^{-1}$ interchanged. (In the following such terms
will be denoted by $q\leftrightarrow q^{-1}$ for simplicity.)
It is easy to verify that (10) and (17) together with (18) lead to (5).
\par
Next let us examine the OPE between $T^{\alpha}(z)$ and $J(w)$:
$$T^{\alpha}(z)J(w)={{1}\over {4}}:J(zq^{\alpha /2})J(zq^{-\alpha /2}):
J(w)~+~q\leftrightarrow q^{-1}.\eqno(19)$$
Since only the singular part is of interest, we have using (12)
$$\eqalign{T^{\alpha}(z)J(w)&\sim {{1}\over {2}}<J(zq^{\alpha
 /2})J(w)>J(zq^{-\alpha /2})~+~q\leftrightarrow q^{-1}\cr
&\sim {{1}\over {2}}{{\kappa}\over {(zq^{\alpha /2}-w)_{q}^{2}}}
J(zq^{-\alpha /2})+q\leftrightarrow q^{-1}.}\eqno(20)$$
Then by expanding the field $J(zq^{-\alpha /2})$ using the
$q$-Taylor's series
\footnote\dag{In (21) we only retain
the first two terms of the expansion as the rest contain the
factor $(zq^{-\alpha /2}-w)_{q}^{2}$ which cancels with the term in the
denominator of (20).}: (see ref.[7])
$$J(zq^{-\alpha /2})=J(wq^{-\alpha -1}) + (zq^{-\alpha
/2}-wq^{-\alpha -1})\partial _{w}^{q}J(wq^{-\alpha}) + ... \eqno(21)$$
the above OPE reduces to
$$T^{\alpha}(z)J(w)\sim{{\kappa}\over {2}}\{{{J(wq^{-\alpha -1})}\over
{(zq^{\alpha
/2}-w)_{q}^{2}}}~+~{{q^{-\alpha}\partial_{w}^{q}J(wq^{-\alpha})}\over
{(zq^{\alpha /2}-wq)}}\}~+~ q\leftrightarrow q^{-1}\eqno(22)$$
which is the singular part of the $q$-OPE between $T^{\alpha}(z)$ and
$J(w)$.
\par
It is again instructive to compare the commutator between
$L_{m}^{\alpha}$ and $a_n$ obtained from the $q$-OPE above with
that evaluated directly from the commutation relations. To this end
we have after a short computation
$$\eqalign{[L_{m}^{\alpha},a_n] &= \oint_{C}{{dw}\over {2\pi i}}
\oint_{C_P}{{dz}\over {2\pi i}}z^{m+1}w^n T^{\alpha}(z)J(w)\cr
&= -\kappa~{\rm cosh}(\alpha (2n+m)\epsilon /2)[n]a_{m+n}}\eqno(23)$$
which is precisely what one would obtain if the bracket was
evaluated directly from the definition (5) and the relations (11).
\par
With the $q$-OPE between $T^{\alpha}(z)$ and $J(w)$ so obtained we
can go on to compute the $q$-OPE between $T^{\alpha}(z)$ and
$T^{\beta}(w)$. Indeed, by writing
$$T^{\alpha}(z)T^{\beta}(w) = \lim_{w' \to w}
{{1}\over {4}}T^{\alpha}(z):J(wq^{\beta /2})J(w'q^{-\beta /2}):~~+~~
q\leftrightarrow q^{-1}\eqno(24)$$
and using (22) we have after a lengthy calculation
$$\eqalign{T^{\alpha}(z)T^{\beta}(w) \sim {{\kappa}
\over {2(q-q^{-1})w}}
\{ &{{T^{\alpha +\beta +1}(wq^{(\alpha + 1)/2})}\over
{(zq^{-(\alpha -\beta)/2}-wq^{\beta +1})}}
+{{T^{-\alpha +\beta -1}(wq^{(\alpha + 1)/2})}\over
{(zq^{-(\alpha +\beta)/2}-wq^{-\beta +1})}}\cr
-&{{T^{-\alpha -\beta +1}(wq^{(\alpha - 1)/2})}\over
{(zq^{-(\alpha -\beta)/2}-wq^{\beta -1})}}
-{{T^{\alpha -\beta -1}(wq^{(\alpha - 1)/2})}\over
{(zq^{-(\alpha +\beta)/2}-wq^{-\beta -1})}}\rbrace\cr
+{{\kappa}^2\over {4(q-q^{-1})w^3}}\{
&{{1}\over{(q^{\alpha +\beta /2 +1}-q^{-\beta /2})_{q}^{2}}}
{{1}\over{(zq^{-(\alpha -\beta) /2 }-wq^{\beta +1})}}\cr
+&{{1}\over{(q^{\alpha -\beta /2 +1}-q^{\beta /2})_{q}^{2}}}
{{1}\over{(zq^{-(\alpha +\beta) /2 }-wq^{-\beta +1})}}\cr
-&{{1}\over{(q^{\alpha +\beta /2 -1}-q^{-\beta /2})_{q}^{2}}}
{{1}\over{(zq^{-(\alpha -\beta) /2 }-wq^{\beta -1})}}\cr
-&{{1}\over{(q^{\alpha -\beta /2 -1}-q^{\beta /2})_{q}^{2}}}
{{1}\over{(zq^{-(\alpha +\beta) /2 }-wq^{-\beta -1})}}\rbrace\cr
+&{~~~~~~~~~~~}q\leftrightarrow q^{-1}}\eqno(25)$$
where the terms in the first bracket correspond to the operator
part while those in the second are the anomaly terms. From this
$q$-OPE we can obtain the $q$-Virasoro algebra by evaluating the
integrals in
$$[L_{m}^{\alpha},L_{n}^{\beta}] = \oint_{C}{{dw}\over {2\pi i}}
\oint_{C_P}{{dz}\over {2\pi i}}z^{m+1}w^{n+1} T^{\alpha}(z)
T^{\beta}(w).\eqno(26)$$
For the operator part the terms are identical to those of (7a) but
central term now yields
$$-{{1}\over{16\epsilon ^{2}}}({C'}_{m+1}^{\alpha ,\beta}+
{C'}_{m+1}^{\alpha ,-\beta})\delta_{m+n,0},\eqno(27)$$
where
$$\eqalign{{C'}_{m}^{\alpha , \beta} = &{~}{{{\rm sinh}({{\alpha
+\beta + 2}\over {2}} m\epsilon)}\over{{{\rm sinh}({{\alpha +\beta +
2}\over {2}} \epsilon})}}
-2{\rm cosh}(m\epsilon - \epsilon){{{\rm sinh}({{\alpha +\beta
}\over {2}} m\epsilon)}\over{{{\rm sinh}({{\alpha +\beta }\over {2}}
\epsilon })}}\cr
&+ {{{\rm sinh}({{\alpha +\beta
- 2}\over {2}} m\epsilon)}\over{{{\rm sinh}({{\alpha +\beta -
2}\over {2}} \epsilon })}}.}
\eqno(28)$$
It is interesting to note that in the limit $q\to 1$
(or $\epsilon \to 0$) the central term reduces to
$${{1}\over {12}}m(m-1)(m+1)\delta_{m+n,0}$$
which is usual central term that one has for the Virasoro algebra.
\vfil\eject
{\bf REFERENCES}
\ref 1 Faddev L D 1982 {\it Les Houches Lectures} (Elsevier,Amsterdam, 1984)
\spa {~} Kulish P P and Sklyanin E K  1982 {\it Lecture Notes in Physics Vol.
151} (Springer, Berlin)
\ref 2 Yang C N 1967 {\it Phys. Rev. Lett.} {\bf 19} 1312
\spa {~} Baxter R J 1982 {\it Exactly Solved Models in Statistical
Mechanics} (Ney York: Academic)
\ref 3 Zamolodchikov A and Zamolodchikov Ab 1979 {\it Ann. Phys.}
{\bf 120} 252;
\spa {~} de Vega H 1989 {\it Int. J. Mod Phys.} {\bf 4} 2371
\ref 4 Moore G and Seiberg N 1989 {\it Nucl. Phys.}B {\bf313} 16;
1989 {\it Commun. Math. Phys.} {\bf123} 177
\spa {~} Alvarez-Gaum\'e L, Gomez C and Sierra G 1989 {\it Phys. Lett.}
{\bf 220B} 142
\ref 5 V.G. Drinfeld, Proc. Intern. Congress of Mathematicians
(Berkley 1986) Vol. 1 pg 798;
\ref 6 Curtright T and Zachos C 1990 {\it Phys. Lett.}{\bf B243} 237
\ref 7 Aizawa N and Sato H 1991 {\it Phys. Lett.}{\bf B256} 185
\ref 8 Chaichian M and Pre${\rm {\check s}}$najder P 1992 {\it Phys. Lett.}
{\bf B277} 109
\ref 9 Kemmoku R and Saito S {\sl `Discretisation of Virasoro
Algebra'}, Tokyo Metropolitan University preprint TMUP-HEL-9303,
March 1993.
\ref {10} Celeghini E, Giachetti R, Sorace E and Tarlini M 1990 {\it
J. Math Phys.}{\bf 31} 2548
\ref {11} Macfarlane A J 1989 {\it J. Phys.}{\bf A22} 4581;
\spa {~}Biedenharn L C 1989 {\it J. Phys.}{\bf A22} L873
\ref {12} Oh C H and Singh K 1992 {\it J. Phys.}{\bf A25} L149
\ref {13}  L\"ust  D and Theisen S 1989 {\it Lecture Notes in Physics} Vol
346 (Berlin: Springer)
\vfil\bye